\documentclass[aps,prb,showpacs,twocolumn,citeautoscript,superscriptaddress]{revtex4}
\usepackage{amsmath}
\usepackage{graphicx}
\usepackage{dcolumn}
\usepackage{float}

\begin{document}
\title{Itinerant and local-moment magnetism in EuCr$_{2}$As$_{2}$ single crystals}
\author{U. B. Paramanik}
\affiliation{Department of Physics, Indian Institute of Technology, Kanpur 208016, India}
\author{R. Prasad}
\affiliation{Department of Physics, Indian Institute of Technology, Kanpur 208016, India}
\author{C. Geibel}
\affiliation {Max-Planck Institute for Chemical Physics of Solids, 01187 Dresden, Germany}
\author{Z. Hossain}
\email{zakir@iitk.ac.in}
\affiliation{Department of Physics, Indian Institute of Technology, Kanpur 208016, India}
\affiliation {Max-Planck Institute for Chemical Physics of Solids, 01187 Dresden, Germany}
\date{\today}

\begin{abstract}

We report on the crystal structure, physical properties and electronic structure calculations for the ternary pnictide compound EuCr$_{2}$As$_{2}$. X-ray diffraction studies confirmed that EuCr$_{2}$As$_{2}$ crystalizes in the ThCr$_{2}$Si$_{2}$-type tetragonal structure (space group \textit{I4/mmm}). The Eu-ions are in a stable divalent state in this compound. Eu moments in EuCr$_{2}$As$_{2}$ order magnetically below $T_m$ = 21~K. A sharp increase in the magnetic susceptibility below $T_m$ and the positive value of the paramagnetic Curie temperature obtained from the Curie-Weiss fit suggest dominant ferromagnetic interactions. The heat capacity exhibits a sharp $\lambda$-shape anomaly at $T_m$, confirming the bulk nature of the magnetic transition. The extracted magnetic entropy at the magnetic transition temperature is consistent with the theoretical value $Rln(2S+1)$ for $S$ = 7/2 of the Eu$^{2+}$ ion. The temperature dependence of the electrical resistivity $\rho(T)$ shows metallic behavior along with an anomaly at 21~K. In addition, we observe a reasonably large negative magnetoresistance ($\sim$ -24\%) at lower temperature. Electronic structure calculations for EuCr$_{2}$As$_{2}$ reveal a moderately high density of states of Cr-3$d$ orbitals at the Fermi energy, indicating that the nonmagnetic state of Cr is unstable against magnetic order. Our density functional calculations for EuCr$_{2}$As$_{2}$ predict a G-type AFM order in the Cr sublattice. The electronic structure calculations suggest a weak interlayer coupling of the Eu-moments.

\end{abstract}


\maketitle

\section{INTRODUCTION}

The layered pnictide intermetallic compounds RT$_{2}$Pn$_{2}$ (R = rare-earth elements, T = transition metal; Pn = pnictide) with ThCr$_{2}$Si$_{2}$-type tetragonal structure (space group \textit{I4/mmm}) exhibit a rich variety of transport and magnetic properties. These compounds consist of alternate `T-Pn' layers and `R' layers stacked along the $c$ axis. Following the exploration of these materials over the last 20 years, recently, the discovery of high temperature superconductivity (SC) in the doped AFe$_{2}$As$_{2}$ (A = divalent alkaline metal or rare-earth metal) has generated a new wave of investigations in search of new compounds in this class, which exhibit interesting magnetic and superconducting properties. The Fe atoms in these materials undergo a spin-density-wave (SDW) antiferromagnetic (AFM) transition below 200~K. Upon doping or under application of external pressure, the Fe AFM ordering weakens and SC emerges.\cite{Rotter, Sasmal, Sefat, Jeevan, Miclea}

Europium is among the few special rare-earth elements having two stable valence configurations: Eu$^{2+}$ ($J = S$ = 7/2) and Eu$^{3+}$ ($J$ = 0); Eu$^{2+}$ bears a strong magnetic moment ($\sim$7.0 $\mu_{B}$) whereas Eu$^{3+}$ does not carry any moment. In a few cases a mixed-valence state of Eu is also observed, for example, in EuNi$_{2}$P$_{2}$ and EuCu$_{2}$Si$_{2}$ [6,7,8]. EuFe$_{2}$As$_{2}$ is a member of the Fe based ``122'' pnictide family where Eu is divalent. This system undergoes a SDW transition in the Fe sublattice at 190~K accompanied by an AFM ordering of Eu$^{2+}$ moments at 19~K [9]. The interplay between SC and Eu$^{2+}$ magnetism in doped EuFe$_{2}$As$_{2}$ has been extensively studied recently.\cite{Jeevan, Miclea, Jeevan1, Zapf, Anupam, Paramanik} Replacing As by P in EuFe$_{2}$P$_{2}$, no Fe moment has been observed in the system and the divalent Eu moments order ferromagnetically at $T_C$ = 30~K as has been detected by neutron diffraction measurements.\cite{Feng, Ryan} Incommensurate antiferromagnetic structure of Eu$^{2+}$ moments with $T_N$ = 47~K has been found in EuRh$_{2}$As$_{2}$ [15]. While EuCu$_{2}$As$_{2}$ exhibits a delicate balance between FM and AFM ordering,\cite{Sengupta} EuNi$_{2}$As$_{2}$ and EuCo$_{2}$As$_{2}$ order antiferromagnetically.\cite{Bauer, Ballinger} Briefly, the pnictide compounds of this structure class show a variety of novel and interesting behaviors.

We synthesized a new isostructural compound, EuCr$_{2}$As$_{2}$. Thia compound crystalizes in the ThCr$_{2}$Si$_{2}$-type tetragonal structure with space group \textit{I4/mmm}. As shown in Fig. 1, alternating Eu layers and CrAs layers are stacked along the $c$ axis where Cr atoms form a square planar lattice in the CrAs layer, similar to the AFe$_{2}$As$_{2}$. Recently, Singh et al. have investigated the closely related compound BaCr$_{2}$As$_{2}$ [19]. A combined study of physical properties and electronic structure calculations demonstrate that BaCr$_{2}$As$_{2}$ is a metal with itinerant antiferromagnetism, similarly to the parent phases of Fe-based superconductors but with slightly different magnetic structure. Neutron diffraction measurements on BaFe$_{2-x}$Cr$_{x}$As$_{2}$ crystals reveal that the Cr doping in BaFe$_{2}$As$_{2}$ leads to suppression of the Fe SDW transition but the superconductivity (as usually observed in case of  other transition metal doping) is prevented by a new competing magnetic order of G-type antiferromagnetism which becomes the dominant magnetic ground state for $x$ $>$ 0.3.\cite{Athena, Marty} BaCr$_{2}$As$_{2}$ shows stronger transition metal-pnictogen covalency than the Fe compounds,\cite{Singh} and in that respect is more similar to the widely studied compound BaMn$_{2}$As$_{2}$. BaMn$_{2}$As$_{2}$ has been characterized as a small band-gap semiconductor with G-type AFM ordering of Mn moments at $T_N$ = 625~K [22,23]. This material becomes metallic by partial substitution of Ba by K or by applied pressure on the parent compound.\cite{KBaMn, PBaMn, PKBaMn} In contrast to BaCr$_{2}$As$_{2}$ and BaMn$_{2}$As$_{2}$, both having tetragonal crystal structure, EuMn$_{2}$As$_{2}$ forms in hexagonal crystal structure\cite{Ruhel} whereas EuCr$_{2}$As$_{2}$ is found to be tetragonal. Very recently, the closely related compounds LnOCrAs (Ln = La, Ce, Pr,and Nd) possessing similar CrAs layers as in BaCr$_{2}$As$_{2}$ have been synthesized by Park et al.\cite{Hosono} These compounds are isostructural (ZrCuSiAs-type structure with the space group \textit{P4/nmm}) to that of LnOFeAs, which are the parent compounds of Fe-based high $T_c$ superconductors. Powder neutron diffraction measurements at room temperature reveal that Cr$^{2+}$ ions in LaOCrAs bear a large itinerant moment of 1.57~$\mu_{B}$ pointing along the $c$ axis which undergo a G-type AFM ordering. The N\'{e}el temperature $T_N$ has been estimated to be in between 300-550~K. Therefore, the related materials possessing CrAs layers are highly enthralling with regard to the physical properties when the AFM ordering is suppressed by doping.

Here we report on the crystal structure, physical properties and electronic structure calculations of EuCr$_{2}$As$_{2}$. Our combined experimental investigations and density functional studies show that Eu-ions are in a divalent state and the Eu$^{2+}$ local moments order magnetically at $T_m$ = 21~K. $M(T)$ and $M(H)$ data suggest competing FM and AFM interactions since the $M(T)$ curves look like that of a ferromagnet while the $M(H)$ curves lack the features typically observed in a ferromagnet. A large negative magnetoresistance is found below $T_m$. Density-functional theory-based calculations indicate that the Cr ions
bear itinerant moments and the most stable magnetic state in the Cr sublattice is a G-type AFM order.

\section{METHODS}

The single crystals of EuCr$_{2}$As$_{2}$ were grown using CrAs flux as described by Singh et al.\cite{Singh} The CrAs binary was presynthesized by reacting the mixture of Cr powder and As pieces at 300 $^\circ$C for 10 h, and then at 600 $^\circ$C for 30 h and finally at 900 $^\circ$C for 24 h. A ratio of Eu : CrAs = 1 : 4 was placed in an alumina crucible, and sealed inside a tantalum tube. The assembly was put into a furnace and heated to 1230 $^\circ$C slowly and held there for 13 hours, and then was cooled to 1120 $^\circ$C at a rate of 2 $^\circ$C/h, finally it was furnace-cooled to room temperature. The shiny plate-like EuCr$_{2}$As$_{2}$ crystals were formed in layers, which were cleaved mechanically from the flux. Several plate like single crystals with typical dimension $4\times4\times0.2$ mm$^3$ were obtained. The polycrystalline samples of EuCr$_{2}$As$_{2}$ were prepared using solid state reaction method similar to that of EuFe$_{2}$As$_{2}$ as described in our earlier reports. \cite{Jeevan, Jeevan1, Anupam} Stoichiometric amounts of the starting elements of Eu chips (99.9\%), Cr powder (99.999\%), and As chips (99.999\%) were used for the reaction. The Single crystals and crushed polycrystalline samples were characterized by x-ray diffraction (XRD) with Cu-$K_\alpha$ radiation to determine the single phase nature and crystal structure. Scanning electron microscope (SEM) equipped with energy dispersive x-ray (EDX) analysis was used to check the homogeneity and composition of the samples. The electrical transport properties were measured by standard four probe technique using close cycle refrigerator (Oxford Instruments) and physical property measurement system (PPMS-Quantum design). The $\chi(T)$ = $M(T)$/$H$ and $M(H)$ isotherms were measured using a commercial SQUID magnetometer (MPMS, Quantum-Design). The specific heat was measured by relaxation method in a PPMS-Quantum design.

We have carried out the density-functional band structure calculations using the full potential linear augmented plane wave plus local orbitals (FP-LAPW+lo) method as implemented in the WIEN2k code.\cite{Blaha} The Perdew-Burke-Ernzerhof (PBE) form of the generalized gradient approximation (GGA) was used to calculate the exchange correlation potential.\cite{Perdew} Additionally, to correct the on-site strong Coulomb interaction within the Eu-$4f$ orbitals we have included U on a mean-field level using the GGA+U approximation. No spectroscopy data for EuCr$_{2}$As$_{2}$ are available in the literature, therefore, we have used U = 8 eV, the standard value for an Eu$^{2+}$ ion.\cite{Jeevan, Jeevan1, Li} In addition, the spin orbit coupling is included with the second variational method in the Eu-$4f$ shell. The set of plane-wave expansion $K_{MAX}$ was determined as $R_{MT}$$\times$$K_{MAX}$ equals to 7.0 and the $K$-mesh used was $10\times10\times10$.

\section{RESULTS AND DISCUSSION}

\subsection*{\label{ExpDetails} A. Crystal Structure}

\begin{figure}[htb!]
\includegraphics[width=8.7cm, keepaspectratio]{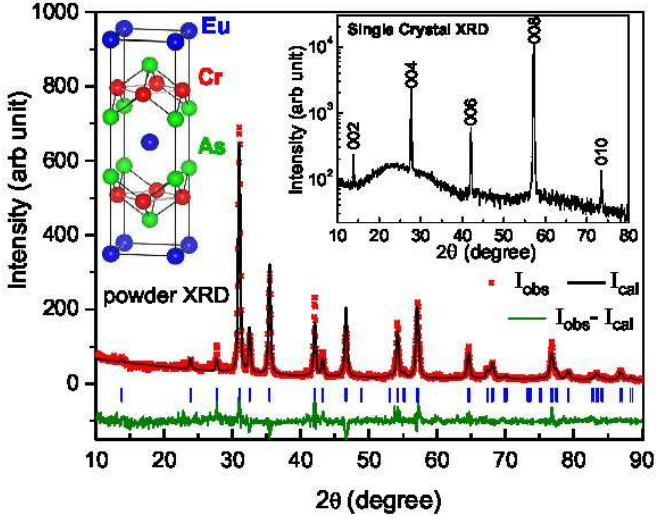}
\caption{\label{fig:XRD} (Color online) (a) The powder x-ray diffraction pattern of EuCr$_{2}$As$_{2}$ recorded at room temperature. The Rietveld refinement fit (solid black line), difference profile (lower solid green line) and positions of Bragg peaks (vertical blue bars) are also shown. Inset: x-ray diffraction pattern for EuCr$_{2}$As$_{2}$ plate-like single crystal.}
\end{figure}

Fig. 1 shows the powder XRD pattern at room temperature for the crushed polycrystalline sample of EuCr$_{2}$As$_{2}$. All the diffraction peaks could be indexed based on the ThCr$_{2}$Si$_{2}$-type structure (space group \textit{I4/mmm}). The crystallographic lattice parameters are listed in Table I. The $c/a$ ratio for EuCr$_{2}$As$_{2}$ is much larger than that of other Eu based transition metal pnictides. A comparison of the structural parameters is shown in Table III.  An increased $c/a$ ratio has also been observed in the homologous compound BaCr$_{2}$As$_{2}$ ($a = 3.96 $~{\AA} ~and~ $c = 13.632 $~{\AA}) [19] as compared to other transition metal compounds BaT$_{2}$As$_{2}$. The inset of Fig. 1 shows the x-ray diffraction pattern for a EuCr$_{2}$As$_{2}$ single crystal. Only the (00l) diffraction peaks are observed, confirming that the crystallographic $c$ axis is perpendicular to the plane of the plate-like single crystals. From the EDX analysis, the single phase nature of the sample is manifested with obtained atomic ratio of Eu : Cr : As as 20.8 : 38.3 : 40.9.

\begin{table}
\caption{\label{tab:XRD} Crystallographic parameters obtained from the structural Rietveld refinement of powder XRD data of EuCr$_{2}$As$_{2}$. The
refinement quality parameter $\chi^{2}$ = 1.62.}
\begin{ruledtabular}
\begin{tabular}{llll}
Structure &\multicolumn{3}{l} {ThCr$_{2}$Si$_{2}$-type Tetragonal} \\
Space group & \textit {I4/mmm} \\
\multicolumn{2}{l}{Lattice  parameters} \\
 \hspace{1cm} $a$ (\AA)  & 3.893(2)  \\
 \hspace{1cm} $c$ (\AA)  & 12.872(2)  \\
 \hspace{1cm} $V_{cell}$ (\AA$^3$)& 195.08(1) \\

Refined Atomic Coordinates\\
\hspace{0.5cm}Atom \hspace{0.5cm} Wyckoff & x &y ~&~~ z  \\
\hline
\hspace{0.5cm} Eu \hspace{1.0cm}~2a & 0 & 0 ~&~~ 0 \\
\hspace{0.5cm} Cr \hspace{1.0cm}~4d & 0 & 0.5 ~&~~ 0.25 \\
\hspace{0.5cm} As \hspace{1.0cm}~4e & 0 & 0 ~&~~ 0.363 \\

\end{tabular}
\end{ruledtabular}
\end{table}

\subsection*{\label{ExpDetails} B. Magnetic susceptibility and isothermal magnetization}

\begin{figure}
\includegraphics[width=8.7cm, keepaspectratio]{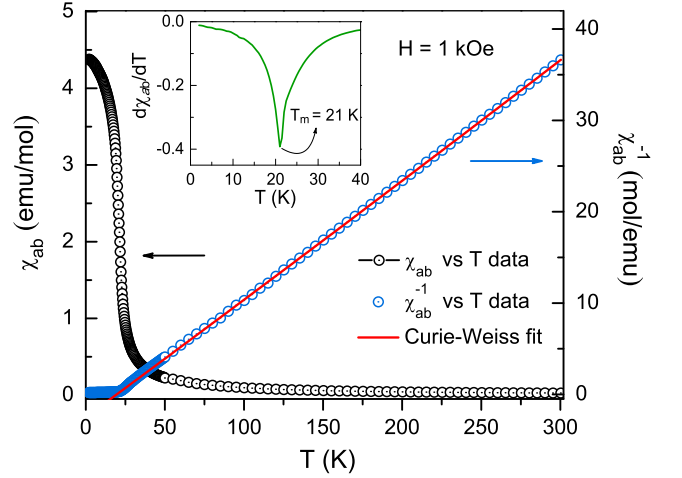}
\caption{\label{fig:MT} (Color online) Temperature dependence of magnetic susceptibility $\chi$$_{ab}$ for EuCr$_{2}$As$_{2}$ with the applied magnetic field H = 1~kOe. The solid line represents the fit to the Curie-Weiss behavior.}
\end{figure}

Fig.~\ref{fig:MT} shows the temperature dependence of the magnetic susceptibility $\chi$$_{ab}(T)$ for EuCr$_{2}$As$_{2}$ with the applied magnetic field H = 1 kOe along the crystallographic ab-plane (H$\parallel$$ab$). There is a sharp increase in $\chi$$_{ab}(T)$ below 21~K which tend to saturate at lower temperature as in the case of  a ferromagnetic order. At high temperature $\chi$$_{ab}(T)$ follows the modified Curie-Weiss behavior, $\chi(T) = \chi_{0} + C/(T - \theta_{P})$ where $\chi_{0}$ is the temperature-independent term of the susceptibility, C is the Curie constant and $\theta$$_{P}$ is the Weiss temperature. The fitting of inverse susceptibility data by the Curie-Weiss behavior in the temperature range 50-300~K (shown by the solid line) yields the effective paramagnetic moment $\mu_{eff}$ = 7.95~$\mu_{B}$ and $\theta$$_{P}$ = 19~K. Similar fit for $\chi$$_{c}(T)$ data (not shown here) yields the effective paramagnetic moment $\mu_{eff}$ = 8.27~$\mu_{B}$ and $\theta$$_{P}$ = 22~K. For both $\chi$$_{ab}$ and $\chi$$_{c}$, the effective paramagnetic moments are close to the theoretical value of g$\sqrt{S(S+1)}\mu_{B}$ = 7.94 $\mu_{B}$ for free Eu$^{2+}$ moments ($S$ = 7/2, $L$ = 0). The positive values of the paramagnetic Curie temperature $\theta_{P}$ obtained from the fit suggest predominantly ferromagnetic exchange interactions between the Eu$^{2+}$ moments.

\begin{figure}
\includegraphics[width=4.4cm, keepaspectratio]{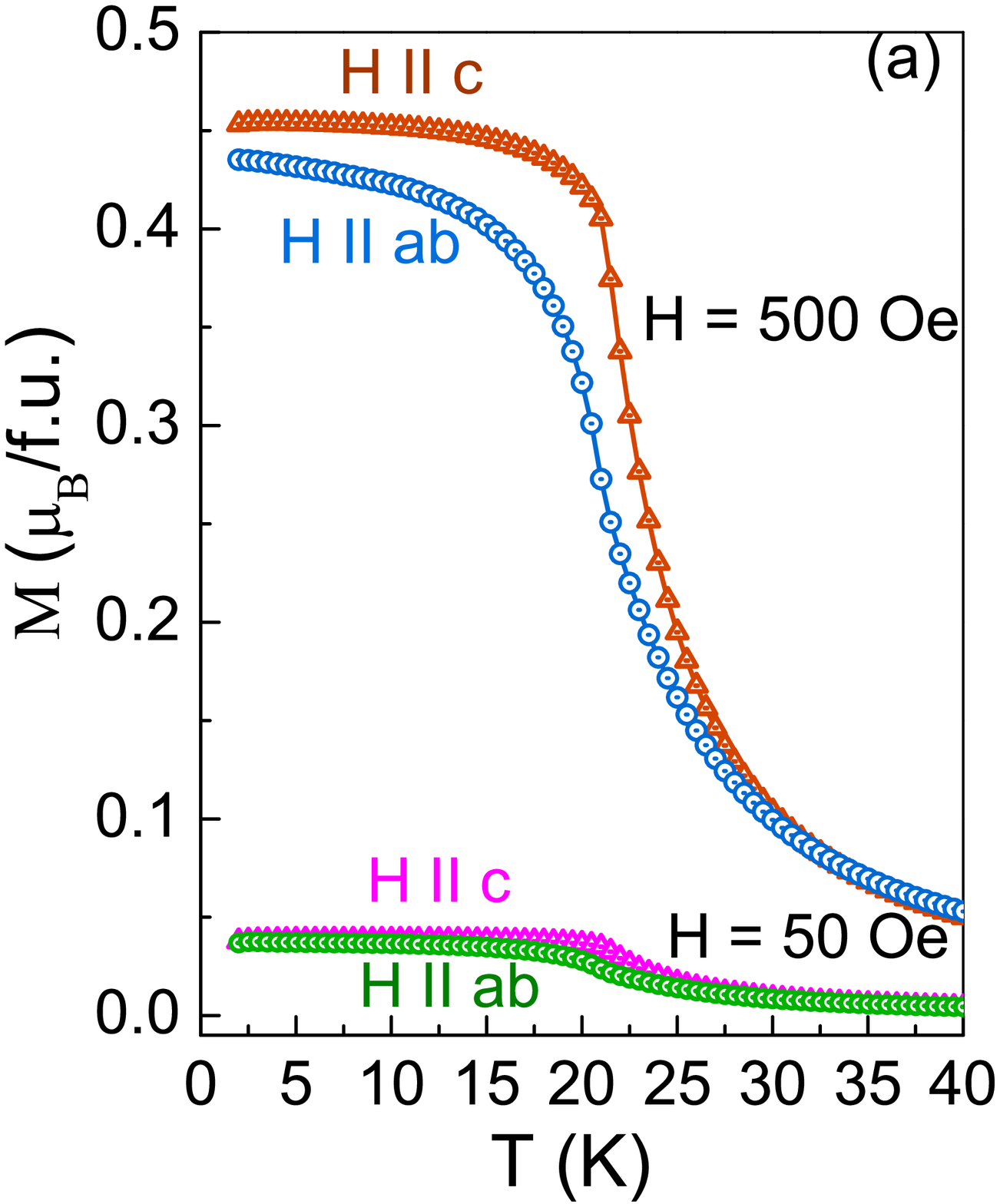}
\includegraphics[width=4.1cm, keepaspectratio]{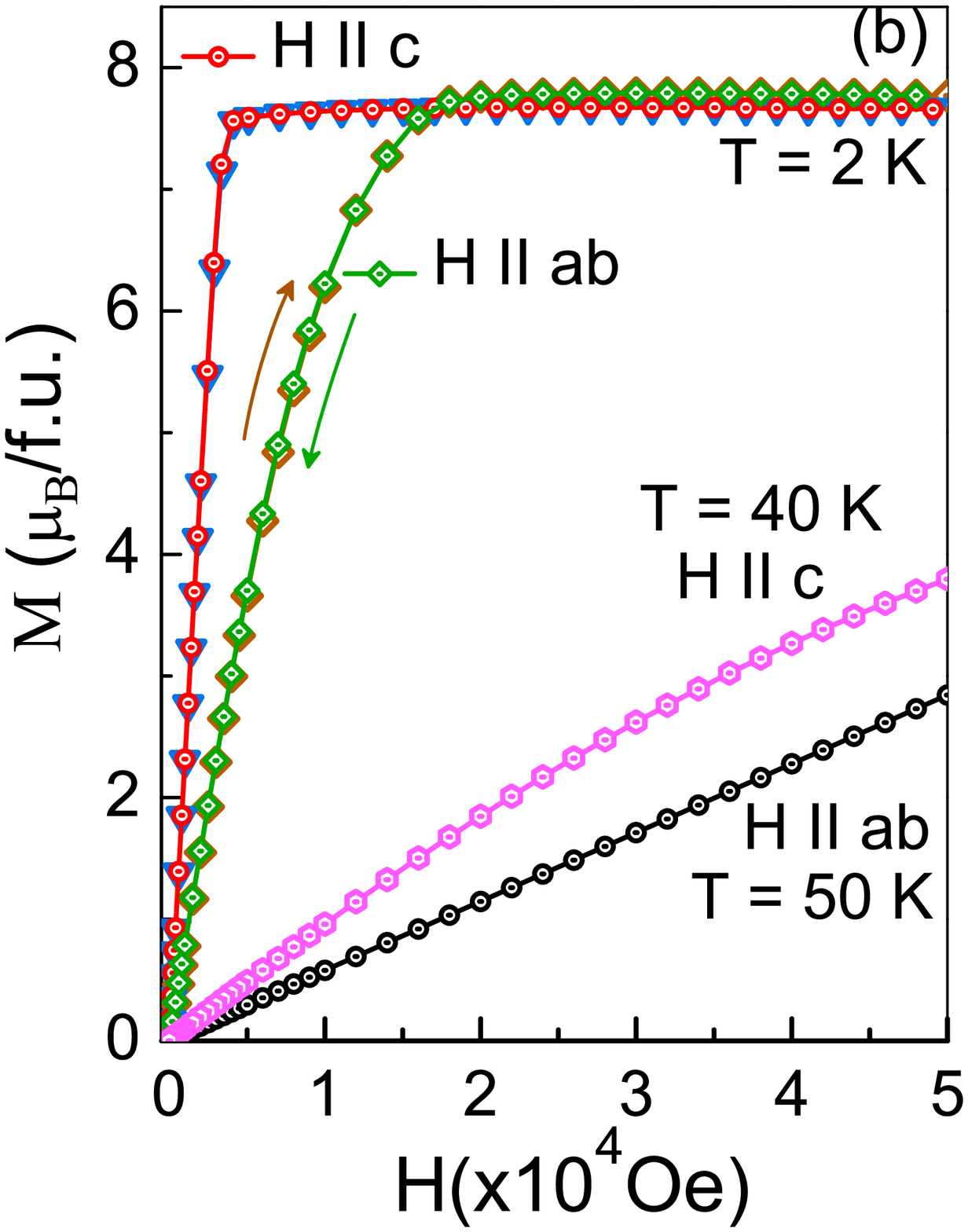}
\caption{\label{fig:MTH} (Color online) (a)  Temperature dependence of magnetization M of EuCr$_{2}$As$_{2}$ single crystal under applied field of 50 Oe and 500 Oe with H in the ab plane (H$\parallel$$ab$) and parallel to the crystallographic $c$ axis (H$\parallel$$c$).  All the data shown are in zero field cooled (ZFC) condition. (b) Isothermal magnetization M of EuCr$_{2}$As$_{2}$ single crystal with H$\parallel$$ab$ and H$\parallel$$c$. M-H data were corrected for the demagnetization effect, taken on a plate-like sample.}
\end{figure}

Fig. 3(a) represents the magnetization M (T) in two different orientations of magnetic field, i.e. H$\parallel$$ab$ and H$\parallel$$c$. At high temperature M(T) is almost isotropic, as normally observed for a stable divalent Eu state. Since it bears a spin only ($J = S$ = 7/2) moment, one expects a negligible anisotropy. However, a significant anisotropy in the magnetization is developed below 25~K  (M$_{H\parallel c}$/M$_{H\parallel ab}$ $\approx$ 1.5 at 21 K), suggesting an anisotropic magnetic interaction. The rapid increase of $M(T)$ below 21~K gives an impression that the magnetic order is either ferromagnetic in nature or it has a strong ferromagnetic component. To gain further insight on the nature of the magnetic order we have carried out isothermal magnetization measurements with varying magnetic fields at fixed temperatures [Fig. 3(b)]. At temperature $T$ = 2~K, the magnetization for H$\parallel$$c$ saturates more rapidly as the magnetic field is increased from H = 0 to 4.1 kOe. For H$\parallel$$ab$, the magnetization saturates at much higher field (18~kOe). A sizable magnetic field is required to achieve the saturation of magnetization for both H$\parallel$$ab$ and H$\parallel$$c$. We do not observe any hysteresis in the M(H) curve at 2~K. It is known that a good quality single crystal of a ferromagnet with small anisotropy and thus small domain wall energy may not always exhibit remanent magnetization.\cite{Givord} Thus, one cannot rule out the possibility of a FM state of Eu$^{2+}$ moments in EuCr$_{2}$As$_{2}$. However, combining the experimental results with the electronic structure calculations (to be discussed below) where we get an antiferromagnetic ground state of the interlayer Eu moments with very weak interlayer coupling, it is possible that the dominant nearest-neighbor (i.e. intralayer) Eu-Eu interaction is FM as indicated by a positive value of $\theta$$_{P}$, but there could be a weak or frustrated AFM coupling between the layers. In this connection, we may recall other homologous compounds EuCu$_{2}$As$_{2}$ or EuFe$_{2}$As$_{2}$, wherein the intralayer Eu-Eu interaction has been established to be ferromagnetic, while the interlayer antiferromagnetic coupling is very weak.\cite{Sengupta, Kasinathan, Shuai, Xiao} If we consider the interlayer Eu-Eu antiferromagnetic coupling to be very weak in EuCr$_{2}$As$_{2}$, then, a small external magnetic field as employed in our magnetic measurements can reorient the spin arrangement along the field direction at the onset of magnetic ordering. The saturated magnetization at 2~K is determined to be $\sim$7.78~$\mu_{B}$/f.u and $\sim$7.66~$\mu_{B}$/f.u for H$\parallel$$ab$ and H$\parallel$$c$ respectively, implying that the system is nearly isotropic. The measured saturated magnetization for both the directions are more than that expected for parallelly aligned Eu$^{2+}$ moments (gS = 7.0~$\mu_{B}$/f.u. with g = 2, $S$ = 7/2), indicating that the Cr ions carry an itinerant moment and are contributing to the total magnetization. The electronic structure calculations on EuCr$_{2}$As$_{2}$ (to be discussed below) also suggest that Cr carries an itinerant moment and the most stable magnetic structure in the Cr sublattice is a G-type AFM order. The homologous compound BaCr$_{2}$As$_{2}$ has been proposed to be a metal with itinerant antiferromagnetism.\cite{Singh} In fact, neutron diffraction measurements on BaFe$_{2-x}$Cr$_{x}$As$_{2}$ crystals reveal that for $x$$>$0.3, the magnetic ground state is consistent with G-type AFM order.\cite{Marty} It also suggests that the Cr magnetic ordering could be well above room temperature in the BaCr$_{2}$As$_{2}$ parent compound similar to BaMn$_{2}$As$_{2}$ [23] which also exhibits a G-type AFM ordering of Mn moments at $T_N$ = 625~K. Moreover, the closely related compounds LnOCrAs possessing similar CrAs layers, in which, Cr ions bear a large itinerant moment of 1.57 $\mu_{B}$ and undergo a G-type AFM ordering with N\'{e}el temperature in between 300--550~K.\cite{Hosono} Similar magnetic ordering of Cr in EuCr$_{2}$As$_{2}$ is also possible at higher temperature.

\subsection*{\label{ExpDetails} C. Specific heat}

Fig. 4 shows the plots of temperature dependence of heat capacity $C_{P}(T)$ of the EuCr$_{2}$As$_{2}$ singe-crystal and that of the reference compound BaCr$_{2}$As$_{2}$ taken from ref. 19. The $C_{P}(T)$ of EuCr$_{2}$As$_{2}$ exhibits a sharp $\lambda$-type anomaly due to the magnetic transition at $T_m$ = 21~K, indicating that the magnetic transition is of second-order. The anomaly in $C_{P}(T)$ remains undisturbed under applied field of 500~Oe but with increasing field up to 5~kOe the anomaly is reduced significantly in height and considerably broadened suggesting a field induced change of the nature of the magnetic transition, presumably a field stabilized ferromagnetic order. The magnetic anomaly in the $C_{P}(T)$ makes it difficult to fit the data at lower temperature to extract the electronic specific-heat coefficient ($\gamma$). The measured value of $C_{P}(T)/T$ at 2~K is $\approx$  225~mJ/mol\,K$^{2}$, but the estimation of $\gamma$ from this value is not reliable as there are magnon contribution from the nearby magnetic ordering of Eu moments. A large $C_{P}(T)/T$ ($\approx$  250~mJ/mol\,K$^{2}$) at 2~K was also observed in ferromagnetically ordered EuFe$_{2}$P$_{2}$ [13].

\begin{figure}[htb!]
\includegraphics[width=8.8cm, keepaspectratio]{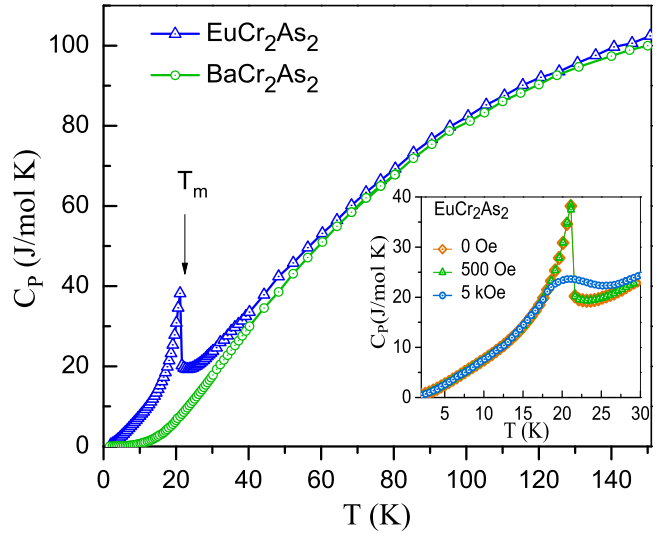}
\caption{\label{fig:cp} (Color online) Temperature dependence of the specific heat $C_{P}$ of EuCr$_{2}$As$_{2}$ single crystal and that of the reference compound BaCr$_{2}$As$_{2}$ taken from ref. 19. The lower inset shows the $C_{P}$ of EuCr$_{2}$As$_{2}$ at zero field and under external fields of 500~Oe and 5~kOe.}
\end{figure}

\begin{figure}[htb!]
\includegraphics[width=8.5cm, keepaspectratio]{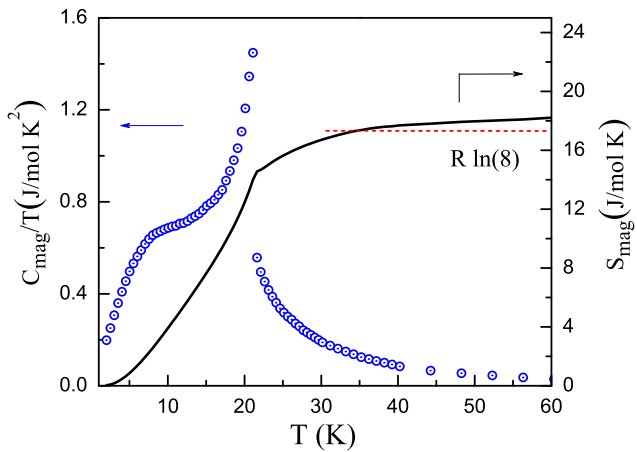}
\caption{\label{fig:sm} (Color online) $C_{mag}/T$ vs T of EuCr$_{2}$As$_{2}$ and the calculated magnetic entropy $S_{mag}$ vs $T$ shown by the solid line.}
\end{figure}

The magnetic part of heat capacity $C_{mag}(T)$ was deduced by the usual method of subtracting the heat capacity of BaCr$_{2}$As$_{2}$ from that of EuCr$_{2}$As$_{2}$ after adjusting the renormalization due to different atomic masses of Ba and Eu, although the mass difference is small here.  Based on the mean-field theory,\cite{Blanco} the heat-capacity jump at the magnetic transition is calculated for the two possible magnetic structures: (i) the equal moment (EM) structure where the magnetic moments are the same at all sites and (ii) the amplitude modulated (AM) structure where the magnetic-moment amplitude varies periodically from one site to another. For EM structure, the jump in the heat capacity at the ordering temperature is given by\cite{Blanco}
\begin{equation}
\Delta C_{EM} = 5 \frac{J(J+1)} {(2J^2+2J+1)} R \label{eq:C}
\end{equation}
and for AM structure,
\begin{equation}
\Delta C_{AM} = \frac{10} {3} \frac{J(J+1)} {(2J^2+2J+1)} R \label{eq:C}
\end{equation}
\noindent where $J$ is the total angular momentum and $R$ is the gas constant. By using $J = S$ = 7/2 for divalent Eu, $\Delta$C$_{EM}$ and $\Delta$C$_{AM}$ amounts to 20.15~J/mol\,K and 13.4~J/mol\,K respectively. Our estimated $\Delta$$C$ ($\approx$ 20.25~J/mol\,K) at $T_{m}$ suggests that EuCr$_{2}$As$_{2}$ possesses an EM structure. In addition, a hump appears in the specific heat at $T \sim T_m$/3, which arises naturally within the mean-field theory for a (2$J$+1)-fold degenerate multiplet. The hump is seen in experimental $C_{mag}(T)$ for EuCr$_{2}$As$_{2}$, which is more pronounced in the $C_{mag}/T$ versus $T$ plot (Fig. 5).  The hump in the ordered state is particularly noticeable in compounds containing Eu$^{2+}$ or Gd$^{3+}$ with $S$ = 7/2, and is not visible for lower $S$, e.g., $S$ = 1/2 [36, 37, 38]. The magnetic contribution to the entropy $S_{mag}$ was obtained by integrating the $C_{mag}/T$ versus $T$. The $C_{mag}/T$ data were extrapolated from $T$ = 2~K to $T$ = 0 in order to approximate the missing $C_{mag}/T$ data between 0 and 2~ K. As can be seen from Fig. 5,  the $S_{mag}$ saturates to the expected theoretical value $R ln(2S+1)$ = 17.3~J/mol\,K, where $S$ = 7/2 for Eu$^{2+}$. The magnetic entropy $S_{mag}$ = 14.6~J/mol\,K at $T_{m}$ is 84\% of the theoretical value.

\subsection*{\label{ExpDetails} D. Transport Properties}

\begin{figure}
\includegraphics[width=8.7cm, keepaspectratio]{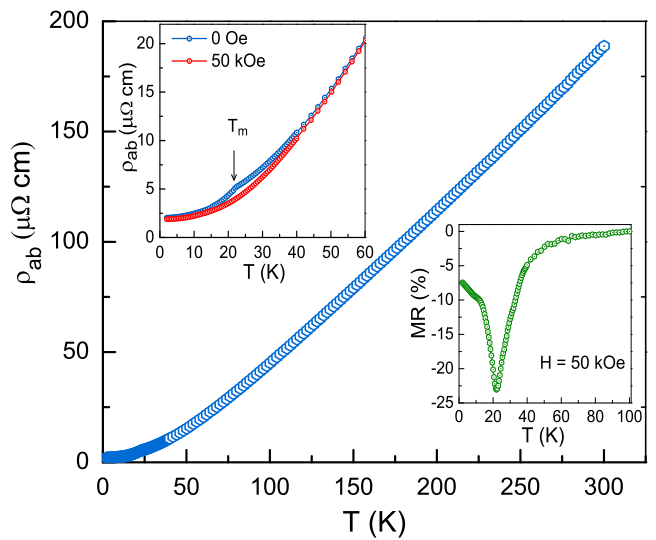}
\caption{\label{fig:Resistivity} (Color online) Temperature dependence of in plane resistivity $\rho$$_{ab}$  for EuCr$_{2}$As$_{2}$ under zero applied magnetic field. The upper inset shows an enlarge view of $\rho$$_{ab}$ under applied magnetic fields of 0 and 50~kOe parallel to the $c$-axis. The lower inset shows the temperature dependence of magnetoresistance for EuCr$_{2}$As$_{2}$.}
\end{figure}

The temperature dependence of in-plane electrical resistivity $\rho_{ab}(T)$ of EuCr$_{2}$As$_{2}$ as shown in Fig.~\ref{fig:Resistivity} exhibits a metallic behavior with residual resistivity $\rho_{ab}$ = 2.0 $\mu\Omega$ cm at 2~K and residual resistivity ratio (RRR) = $\rho_{300\,{K}}/\rho_{2\,{K}} \approx  90$. The  high residual resistivity ratio together with a low residual resistivity confirms the high quality of our crystals. Since the compound is metallic, most likely the magnetic coupling between the Eu spins is mainly mediated by the conduction electrons through indirect Ruderman-Kittel-Kasuya-Yosida (RKKY) interaction. The resistivity data show a kink at the magnetic transition temperature ($T_m$) followed by a rapid decrease in resistivity below $T_m$ due to reduction of spin disorder scattering. Further, we observe significant reduction of the electrical resistivity adjacent to the magnetic ordering temperature on application of magnetic field, leading to a negative magnetoresistance (MR). The MR reaches its maximum value (-24\%) near $T_m$. The MR  [Fig.~\ref{fig:Resistivity}] is defined as [$\rho(H)-\rho(0)]/\rho(0)$, where $\rho(0)$ and $\rho(H)$ are the resistivity measured at zero field and under applied field $H$ = 50~kOe respectively.

\subsection*{\label{ExpDetails} E. Density-functional calculations}
In order to study the electronic and magnetic properties of the compound we start with the calculation of the density of states (DOS) for EuCr$_{2}$As$_{2}$ in the quenched paramagnetic state, that means no spin polarization is allowed on the Cr ions, but, spin polarization is enabled for the Eu ions. Such a study can provide an intimation of magnetic state for the transition metal ions by analyzing the partial density of states (PDOS) at the Fermi level and we can infer whether the magnetic state is favored or not. A similar approach was adopted in prior calculations for EuFe$_{2}$As$_{2}$ [12, 31] and the DOS of which is shown here for comparison purposes. We use the experimental lattice parameters $a = 3.893(2)$~{\AA}, and $c = 12.872(2)$~{\AA} of EuCr$_{2}$As$_{2}$ for the calculations. The internal coordinate of As (z$_{As}$ = 0.361) is determined by force minimization, which is very close to the experimental z$_{As}$ for EuCr$_{2}$As$_{2}$. For the reference compound EuFe$_{2}$As$_{2}$, the experimental lattice parameters were taken from ref. 9.

\begin{figure}[htb!]
\includegraphics[width=8.5cm, keepaspectratio]{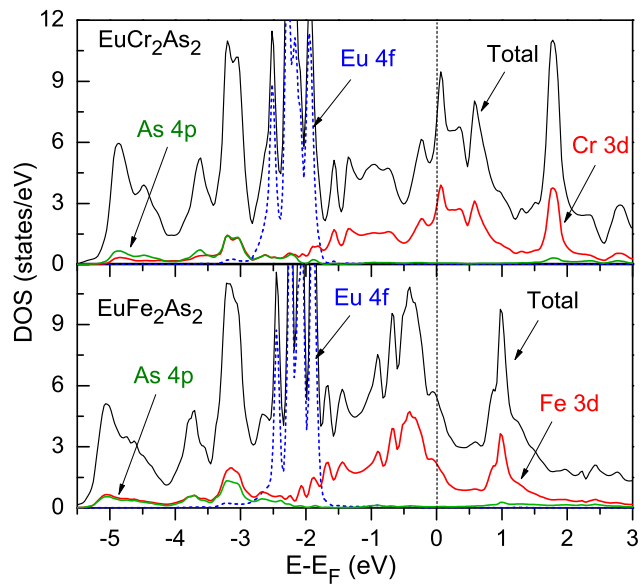}
\caption{\label{fig:dos} (Color online) Total and partial Densities of states (DOS) for EuCr$_{2}$As$_{2}$ and EuFe$_{2}$As$_{2}$  in the NM state in Cr/Fe sublattice and FM interaction between the intralayer Eu spins in the Eu sublattice.}
\end{figure}

The general shape of our density of states for EuCr$_{2}$As$_{2}$ (Fig.~\ref{fig:dos}) is similar to that for EuFe$_{2}$As$_{2}$, but, with a shift of 3$d$ orbitals in the binding energy, which is expected as Cr has two $3d$ electrons less as compared to Fe. The calculated DOS for EuFe$_{2}$As$_{2}$ is very similar to that reported by Li et al.\cite{Li} The Eu $4f$ states for both the compounds are quite localized in between -1.5 to -3~eV, suggesting that the Eu ions are in a stable 2+ valence state. The calculated spin moment for Eu$^{2+}$ is about 6.9 $\mu_{B}$ which is consistent with the experimental value. The rest of the DOS can be divided into two parts: (i) The DOS below -2~eV consists of hybridized Cr $3d$ and As $4p$ orbitals. The $p$-$d$ hybridization between As $4p$ and Cr $3d$ is sizable. (ii) The DOS near the Fermi level ranging from -2 eV to +2 eV is basically composed of the Cr $3d$ orbitals. The Fermi level lies on a steep edge of a peak in the partial density of states of Cr-3$d$ orbitals,  resulting in a relatively large DOS at the Fermi energy. The corresponding value of PDOS at the Fermi level for Cr-$3d$ is N($E_F$) = 3.03 states/eV per Cr atom, which is greater than that of Fe-3$d$ states (2.15 states/eV per Fe atom) in EuFe$_{2}$As$_{2}$ [31]. According to the Stoner criterion, magnetism may occur if $N(E_{F})*I >$ 1, where $I$ is the Stoner exchange-correlation integral.\cite{DJ} We use the  Stoner exchange-correlation integral $I$=~0.38~eV for Cr-3$d$ from the original work of Janak,\cite{Janak} which amounts to $N(E_{F})*I$~=~1.15. Therefore, the non magnetic (NM) state of Cr is unstable against the magnetic order in EuCr$_{2}$As$_{2}$.

\begin{table}
\caption{\label{tab:XRD} Results of Energetic and magnetic properties of EuCr$_{2}$As$_{2}$ for different magnetic states in the Cr sublattice and interlayer AFM coupling in the Eu sublattice. $\Delta E$ (eV) is the total energy difference per formula unit basis (two Cr atoms) with respect to the the non-spin-polarize or non-magnetic state, and $m_{Cr}(\mu_{B})/ m_{Eu}(\mu_{B})$ is the calculated magnetic moment on Cr/Eu. (NM = non-magnetic or non-spin-polarized, FM = ferromagnetic, S-AFM = stripe type AFM order, G-AFM = G-type AFM or checkerboard nearest-neighbor AFM order).}
%\begin{center}
\begin{ruledtabular}
\begin{tabular}{c c c}
Cr-ordering    &  $\Delta$ E(eV) &  $m_{Cr}(\mu_{B})/ m_{Eu}(\mu_{B})$ \\
\hline \\ [0.02ex]
NM  & 0 & 0/6.9  \\ [1ex]
FM   &  -0.136 & 1.28/6.9\\ [1ex]
S-AFM   & -0.064 & 1.69/6.9 \\ [1ex]
G-AFM  & -0.417 & 2.10/6.9 \\ [0.5ex]
\end{tabular}
\end{ruledtabular}
%\end{center}
\end{table}
\begin{figure}
\includegraphics[width=5.5cm, keepaspectratio]{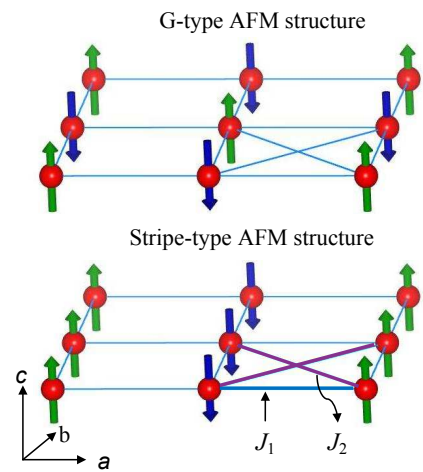}
\caption{\label{fig:afm} (Color online) The top panel shows the G-type (N\'{e}el or checkerboard) AFM structure where nearest-neighbor spins are aligned antiparallel. The bottom panel represents stripe-type AFM ordering along with the definitions of the in-plane exchange constants $J_1$ and $J_2$. }
\end{figure}

\begin{table*}[htb!]
\caption{A comparison of the structural and magnetic parameters of EuCr$_{2}$As$_{2}$ with some isostructural Eu compounds. The lattice parameters $a$ and $c$ are at room temperature, $T_{N}/T_{C}$ are the antiferromagnetic/ferromagnetic ordering temperatures and $\mu_{eff}$ is the effective moment calculated from the Curie-Weiss fit of magnetic susceptibility data.}
\begin{ruledtabular}
\begin{tabular}{c c c c c c}
Compound   & $a$({\AA})    &  $c$({\AA})   &    $T_{N}/T_{C} (K)$  &  $\mu_{eff}$($\mu_{B}$/f.u.)  & Ref.\\
\hline \\ [0.02ex]
EuFe$_{2}$As$_{2}$    &3.907  &12.114  &19  &7.79  &9\\ [1ex]
EuCu$_{2}$As$_{2}$    &4.260  &10.203  &15  &7.90  &16\\ [1ex]
EuNi$_{2}$As$_{2}$    &4.096  &10.029  &14  &7.30    &17,42\\ [1ex]
EuCo$_{2}$As$_{2}$    &3.934  &11.511  &39  &7.40    &18,42\\ [1ex]
EuCr$_{2}$As$_{2}$    &3.893  &12.872  &21  &7.95  &This work\\ [0.5ex]
\end{tabular}
\end{ruledtabular}
\label{comparison}
\end{table*}

To examine the most stable magnetic structure of Cr in EuCr$_{2}$As$_{2}$, we have calculated the total energy for different possible magnetic states, namely, (i) a non-spin polarized calculation (no magnetism on Cr), (ii) FM spin configuration, (iii) stripe-type AFM (similar to that in  EuFe$_{2}$As$_{2}$) and (iv) G-type AFM. The corresponding total energies of different magnetic states together with the calculated moment values are listed in Table II. It is shown that a G-type AFM order in the Cr sublattice is the lowest energy state for EuCr$_{2}$As$_{2}$. The large energy differences between different Cr magnetic configurations suggest that the magnetic ordering temperature of Cr moments should be high. BaCr$_{2}$As$_{2}$ also possess a G-type AFM ground state of Cr itinerant moments as has been reported by Singh et al.\cite{Singh} Neutron diffraction studies on BaFe$_{2-x}$Cr$_{x}$As$_{2}$ show a G-type AFM ground state for $x>$ 0.3 [21]. Recent experimental investigation on the closely related compound LaOCrAs reveal a G-type AFM order of Cr itinerant moments of 1.57 $\mu_{B}$ [28]. Therefore, our calculation of minimum ground state energy for a G-type AFM order of Cr moments in EuCr$_{2}$As$_{2}$ agrees with the magnetic structure of other related compounds. According to the Heisenberg model with nearest-neighbor ($J_1$) and next-nearest-neighbor ($J_2$) spin interactions, the differences in the ordered energies for several collinear commensurate magnetic structures are given by \cite{Johnston}
\begin{equation}
E_{FM} - E_{G-AFM} = 2NS^2(2J_1)
\end{equation}
and
\begin{equation}
E_{G-AFM} - E_{S-AFM} = 2NS^2(2J_2 - J_1)
\end{equation}
\noindent where $N$ is the number of spins $S$.
\noindent The in-plane G-type AFM is favored when $J_1 >$ 0 and $J_1 > 2J_2$. Our calculations yield an antiferromagnetic $J_1$ ($> 0$) and a large negative $J_2$ with $J_2$/$J_1$ = -0.77. The large negative value of $J_2$ implies that this model is probably not reliable for the system. This might be expected in a itinerant magnetic system with long range magnetic order, as has been pointed out by Singh et al. for the BaCr$_{2}$As$_{2}$ system.\cite{Singh}

Experimentally we do not observe any signature of Cr moment ordering up to 300~K. This is not surprising considering the large total energy differences between different magnetic Cr moment configurations, which suggest an ordering temperature well above the maximum temperature of our measurements. Magnetic measurements at higher temperature are needed to corroborate the expected AFM ordering of Cr moments. The calculated magnetic structure of Cr moments can be verified experimentally using neutron or x-ray scattering measurements.

Finally, we discuss the magnetic order in the Eu sublattice. The calculated total energy for the system is found to be minimum when the interlayer Eu spins are antiferromagnetically coupled. Nevertheless, the difference in total energy is very small ($\sim$ 0.0006~eV) whether the interlayer Eu spins are antiferromagnetically coupled or ferromagnetically coupled, implying a rather weak interlayer coupling in the Eu sublattice. So, it is expected that any small external effect (doping, external pressure or external magnetic field) can easily flip the Eu spin from AFM to FM. The P doped EuFe$_{2}$As$_{2}$ system witnesses a similar weak interlayer coupling (0-6~meV), wherein the antiferromagnetic Eu moments arrangement changes from AFM to FM with slight change in doping concentration.\cite{Kasinathan} Furthermore, the homologous system EuFe$_{2}$As$_{2}$ with antiferromagnetic ground state experiences a field-induced spin reorientation to the FM state for an applied field of just 1~T in the $ab$-plane and at 2~T along the $c$-axis,\cite{Shuai, Xiao} which suggests a weak AFM coupling between the interlayer Eu spins. Taking into consideration the relatively large interlayer distance between the Eu layers along the $c$ axis (6.44~{\AA} for EuCr$_{2}$As$_{2}$, and 6.057~{\AA} for EuFe$_{2}$As$_{2}$), the interlayer coupling of Eu spins in EuCr$_{2}$As$_{2}$ is expected to be even lesser.

\section{\label{Conclusions} CONCLUSIONS}

In summary, we have successfully synthesized single and poly crystals of EuCr$_{2}$As$_{2}$ and characterized them using x-ray diffraction, electrical resistivity $\rho(T)$, magnetization and specific heat $C_p(T)$ measurements. The powder XRD data confirm that this compound crystallizes in the body-centered tetragonal structure (space group \textit{I4/mmm}). The $C_p(T)$ and $\rho(T)$ data show anomalies at a temperature $T_m$ = 21~K. While the susceptibility behavior apparently indicates a ferromagnetic order below 21~K, the magnetization data in the ordered state do not show any hysteresis or spontaneous magnetization. Furthermore, the value of $\theta$$_{P}$ obtained from the Curie-Weiss fit in the paramagnetic state is positive and very close to the magnetic transition temperature. These observations indicate that the dominant nearest-neighbor (i.e. intralayer) Eu-Eu interaction is FM but there could be a weak or frustrated AFM coupling between the layers. Also, we do not rule out the possibility of a FM state of Eu$^{2+}$ moments in EuCr$_{2}$As$_{2}$. The measured saturated magnetization for both H$\parallel$$ab$ and H$\parallel$$c$ are larger than the theoretical value of g$S$ = 7.0~$\mu_{B}$ per Eu atom, suggesting that the Cr moments possibly contribute to the observed saturated magnetization values. The $\rho(T)$ data confirm  the metallic state of EuCr$_{2}$As$_{2}$ with a negative magnetoresistance (-24\%) around the magnetic transition. The magnetic entropy $S_{mag}(T)$ at $T_m$ is 84\% of the theoretical value $Rln(2S+1)$ for $S$ = 7/2 of the Eu$^{2+}$ ion and the remaining 16\% is recovered by $\approx$ 34~K. The electronic structure calculations indicate that the Cr ions carry itinerant moment and the most stable magnetic structure in the Cr sublattice is a G-type AFM order. Moreover, the large total energy differences between different magnetic Cr moment configurations suggest an ordering temperature well above the maximum temperature of our measurements. Higher temperature magnetic measurements are needed to observe the expected Cr moment ordering. Density-functional calculations suggest a very weak interlayer coupling between the Eu moments. It would be useful and interesting to determine the magnetic structures of EuCr$_{2}$As$_{2}$ by magnetic neutron or x-ray scattering measurements.

\section*{ACKNOWLEDGEMENTS}

This work has been partially supported by the Council of Scientific and Industrial Research, New Delhi (Grant No. 80(0080)/12/ EMR-II).

\end{document}